# Cation-Dependent Nonadiabaticity in Proton-Coupled Electron Transfer at Electrified Solid-Liquid Interfaces

Tomoaki Kumeda and Ken Sakaushi*

Center for Green Research on Energy and Environmental Materials,

National Institute for Materials Science,

1-1 Namiki, Tsukuba, Ibaraki 305-0044, Japan

*SAKAUSHI.Ken@nims.go.jp

Here we report an observation of cation-dependent nonadiabaticity in a proton-coupled electron transfer electrode process (PCET) during hydrogen evolution at electrified interfaces of Au(111) single-crystal electrode and electrolytes by employing electrochemical kinetic isotope effect analysis at ultra-clean/well-defined systems to investigate quantum-to-classical transition (QCT) in highly accuracy. At low overpotentials, a classical process is dominant independently of the cation environment. However, surprisingly, at high overpotentials, the classical process transforms to a quantum process in the solution containing $K^+$ on the other hand stay classical in the solution containing $Li^+$. Thus, the $K^+$-system shows an anomalous QCT in PCET, which is the inverse direction of the standard QCT under the framework of Brønsted–Evans–Polanyi principle predicting quantum process at low overpotential and classical at high. This anomalous QCT is the manifestation of nonadiabatic PCET and depends on the selection of cations. The nonadiabatic PCET requires the specific hydrogen-bonding structure in the water dimer at electrode surface to emerge nonadiabaticity. $K^+$ was found to be able to break the rigid interfacial water structure at high overpotential to form the appropriate interfacial water configuration to trigger nonadiabatic PCET.

Unveiling the kinetics and mechanism of proton transfer is of important to understand principles in various key energy conversion processes [1−6]. Especially, the proton transfer process at solid-liquid interfaces has been widely studied by experimental and theoretical approaches towards a wide spectrum of electrode processes [7−10]. However, elucidation of the proton transfer in this system is of great challenging amid highly complicated reaction paths exchanging between classical and quantum processes depending on the reaction conditions, such as potential and temperature [11−13].

Pronounced quantum effect is known to emerge for reactions related to hydrogen under appropriate conditions, and govern kinetics and mechanism of processes [14−20]. For example, density functional theory (DFT) based simulations suggest a delocalization of hydrogen adsorbed on metal surfaces due to a quantum effect [21], and this quantum effect is suggested as a reason for the invisible nature of the underpotential-deposited hydrogen [22]. As such, the adsorption of hydrogen is one of the most important steps for electrode processes at solid-liquid interfaces. Potential-induced hydrogen evolution process at solid-liquid interfaces (HER) is known as a fundamental electrocatalysis and also the simplest multielectron-multiproton transfer electrode process. However, the microscopic process of HER is still open to debate. The HER consists of three elementary steps: the adsorption of hydrogen atom (Step 1), the reaction between water and the adsorbed hydrogen (Step 2), and the reaction between the two adsorbed hydrogen atoms (Step 3). In the previous experimental and theoretical studies suggest that a unique quantum phenomenon at electrified solid-liquid interface of nonadaiabaticity indeed emerged in Step 1 [19, 23].

The highly-accurate electrochemical kinetic isotope effect (haEC-KIE) is an advanced experimental method to observe the quantum effect in electrode processes [19, 24−26]. Using this approach, an anomalous potential-dependent transitions between classical and quantum processes (quantum-to-classical transition: QCT) in electrocatalysis was observed in the HER on a polycrystalline Au electrode under a ultraclean conditions [19]. The observed QCT is exotic from the view of Brønsted–Evans–Polanyi (BEP) principle predicting quantum process at low overpotential and classical at high [27−29]. Later, a theoretical study suggested that this anomalous QCT of HER on Au links to diabatic vibronic transition during proton-coupled electron transfer (PCET) [23]. This theoretical framework for the emergence of nonadaiabaticity in the HER electrode process suggests that a proton transfer via proton-donating water molecules with a specific configuration has a pivotal role (Fig. 1).

At the electrified solid-liquid interface, polar molecules such as water are restricted and oriented by electric field and noncovalent interactions with neighboring water or charged ions [30, 31]. Therefore, the effect of water structure at the interface on proton transfer process could be considered as a trigger of the anomalous QCT of HER on Au depending on the potential. Here, we experimentally confirm that the nonadiabaticity of HER on Au(111) surface is tunable by the selection of cations, which was introduce as a descriptor to control water structure at electrified solid-liquid interface. In this study, we investigate the cation-dependent QCT mechanism in the potential-induced HER on Au(111) using haEC-KIE. Our results suggest the role of the interfacial cation and water in the potential-induced proton transfer process.

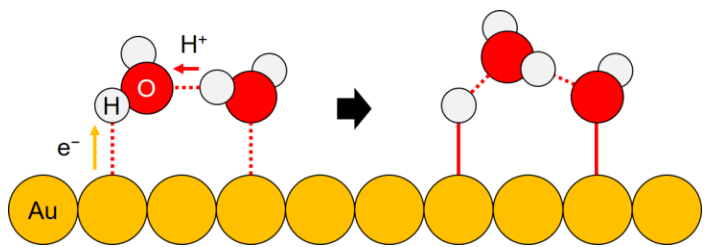

Figure 1: Dimer structure of H-down and H-up water molecules for nonadiabatic proton-coupled electron transfer process on Au(111) in alkaline conditions. Solid and dashed lines respectively represent covalent and non-covalent interactions.

The water structure at solid surfaces has been widely studied by experimental and theoretical approaches [32−35]. According to the previous studies, charged cations and/or the hydration shell can restrict the water structure by noncovalent interactions such as hydrogen bonding or ion-dipole interactions [36−38]. Moreover, the cation is concentrated at the near-surface and blocks electrode surfaces at high overpotential as Koper and his colleagues proposed the promotive or inhibiting effects of cation on the HER activity on an Au surface depending on the potential [39, 40]. By inspiring these previous reports, we hypothesized that cations are keys to tune the theoretically-predicted nonadiabaticity at HER on Au. Thus, here we decided to investigate the HER kinetics on an Au(111) single-crystal electrode in KOH and LiOH solutions.

The well-ordered Au(111) surface used in this study is confirmed by characteristic cyclic voltammetry (CV) (see Fig. S1 of the Supplemental Material). The HER kinetics of the Au(111) electrode in KOH and LiOH using haEC-KIE are discussed based on the diagrams of logarithm of the kinetic HER current density (log*j*) versus overpotential ($\eta$), so-called Tafel plots, in Fig. 2(a) and 2(b). The raw data for the HER polarization curves of Au(111) are shown in the Supplemental Materials (Fig. S2). The Tafel slopes show the kinetic difference between the protonated ($H_2O$) and deuterated ($D_2O$) systems for the HER, which might originate from the difference of vibration frequencies relating to adsorbed hydrogen and deuterium [19]. Furthermore, the HER kinetics is also dependent on the electrolyte cations. Comparing Fig. 2(a) and 2(b), the HER rate on Au(111) in LiOH is slower than that in KOH which is corresponding with previous reports [39, 41]. The origin of the cation-dependent HER kinetics on Au surfaces was suggested to be the affinity between the cation and the HER intermediates, however the detailed mechanism is still under debate.

To analyze the HER kinetics using the haEC-KIE method, we assume the following reaction steps.

$$\text{Au} + \text{H}_2\text{O} \rightarrow \text{Au} \cdots \text{H}_2\text{O}_{ad} \quad \text{(Step I)}$$

$$\text{Au} \cdots \text{H}_2\text{O}_{ad} + e^- \rightarrow \text{Au} - \text{H}_{ad} + \text{OH}^- \quad \text{(Step II)}$$

$$\text{Au} - \text{H}_{ad} + \text{H}_2\text{O} + e^- \rightarrow \text{Au} + \text{H}_2 + \text{OH}^- \quad \text{(Step III)}$$

$$2\text{Au} - \text{H}_{ad} \rightarrow 2\text{Au} + \text{H}_2 \quad \text{(Step III}')$$

The source of hydrogen for the HER is water molecules under alkaline conditions. The first step (Step I) is the adsorption of water on Au sites ($H_2O_{ad}$) with non-covalent interactions (denoted $\cdots$). The second step (Step II) is the electron transfer from the Au substrate and proton transfer from water molecules to form hydrogen atoms adsorbed on Au ($H_{ad}$) with covalent bond (–). The two pathways should be considered in the last step. One is the electron transfer and the proton transfer to $H_{ad}$ forming $H_2$ (Step III). The other is the coupling of the two $H_{ads}$ (Step III'). In previous studies, the rate-determining step (RDS) for the alkaline HER on Au(111) was proposed to be Step II [39, 40, 42]. As shown later, our analysis for the RDS using the transfer coefficient ($\alpha$) also indicated Step II is the RDS. Therefore, we also consider Step II as the RDS in these systems.

The KIE value ($K^{H/D}$) is estimated by comparing the rate constant ratio of $H_2O$ and $D_2O$ systems ($k_H/k_D$), as reported previously (see the Supplemental Material for the details) [19].

Briefly, the KIE value can be obtained by the following equation:

$$K^{H/D} = \frac{[D_2O]j_{0,H}}{[H_2O]j_{0,D}} \exp\left\{\frac{(\alpha_D - \alpha_H)F\eta}{RT}\right\} \tag{1}$$

where $j_0$, $\alpha$, $F$, $\eta$, $R$, and $T$ are the exchange current density, transfer coefficient, Faraday constant, overpotential, gas constant, and temperature, respectively. We can obtain $j_0$ and $\alpha$ from the Tafel plots. Furthermore, $\alpha$ is related with symmetry factor $\beta$ as formulated by Parsons [43]:

$$\alpha = \frac{s}{\nu} + \beta \tag{2}$$

where $s$, and $\nu$ are respectively the electron transfer number before the RDS and stoichiometric number. Fig. 2(c) shows the $\alpha$ vs $\eta$ diagrams for the HER on Au(111) in KOH and LiOH. The $\alpha$ values are measured to be less than 1 in the potential window between $-0.75$ and $-0.5$ V in the both solutions. As $\beta$ takes a value ranging $0 < \beta < 1$, $s$ should be 0 for the systems. From these considerations we obtain $\alpha = \beta$ for the KOH and LiOH conditions. As shown in Fig. 2(d), in the case of the KOH solution, the $K^{H/D}$ value shows more than 1 in the whole potential window. This observation further indicates that proton transfer involves in the RDS. Thus, this series of analysis indicates that the RDS in the KOH system involves the spontaneous transfer of electron and proton therefore is the first concerted proton-electron transfer (CPET), i.e. Step II. Moving to the case of the LiOH, $K^{H/D}$ in this system is 1.0 at $\eta = -0.5$ V and gradually increases then almost saturated at 3.3 at around the overpotential of $-0.75$ V. This observation indicates that the RDS at the low $\eta$ region solely involves electron transfer and this process transform to the first CPET (Step II) at the higher $\eta$ region. Focusing on the potential-dependency of $\beta$, Fig. 2(c) shows that $\beta$ increases along with increase of $\eta$ both in the KOH and LiOH systems as $\alpha = \beta$ in this study. The value of $\beta$ for the KOH and LiOH conditions is almost identical at low $\eta$ region ($< -0.6$ V). However, the $\beta$ for the KOH system increases more rapidly than that of LiOH along with an increase of $\eta$. Because $\beta$ is strongly correlated to the feature of reaction coordinate at a RDS, the different behavior of $\beta$ in KOH and LiOH indicates that the PCET process at the RDS changes more drastically in the KOH condition compared to the LiOH.

In order to unveil the mechanistic origin in the different feature of the electrode processes,

the $K^{H/D}$ values were analyzed. In the both KOH and LiOH conditions, the onset potential for the HER on Au(111) electrodes is approximately $-0.5$ V, as shown in Fig S2. Moreover, at higher overpotentials ($\eta < -0.8$ V), the HER mechanism is suggested to be extremely altered by the cation concentration near the surface [39, 40]. Therefore, $K^{H/D}$ is estimated in the potential range between $-0.75$ V and $-0.5$ V in this study. Fig. 2(d) shows the $K^{H/D}$ vs $\eta$ diagrams for the HER on Au(111) in the KOH and LiOH conditions. $K^{H/D}$ increases along with increase of the overpotential for the both conditions. Interestingly, in the KOH condition, $K^{H/D}$ is 2.0 at $\eta = -0.5$ V and increases drastically to 30 at $-0.75$ V. However, on the other hand, in the LiOH condition, $K^{H/D}$ is 1.0 at $\eta = -0.5$ V and gradually increases then almost saturated at 3.2 at around the overpotential of $-0.75$ V. Under the theoretical framework [44], the maximum $K^{H/D}$ without quantum proton tunneling (QPT) is predicted to be about 10. Therefore, in the KOH condition, we observed the transition potential of the proton transfer mechanism, which shifts from the over-the-barrier path based on the semiclassical transition-state theory (SC-TST) to QPT path around $-0.68$ V. In the LiOH condition, the low $K^{H/D}$ ($K^{H/D} = 1$) indicates solely electron transfer is the RDS at $\eta = -0.5$ V, as mentioned above. And then, the PCET based on the SC-TST dominates in the HER at higher overpotentials than $-0.5$ V in the LiOH condition. To the best of our knowledge, this is the first observation of the cation-dependent quantum-to-classical transition (QCT) of PCET in the potential-induced HER. Furthermore, noteworthy the QCT observed in the HER on Au(111) in the KOH system is anomalous from the view of BEP principle, which predicts that proton tunneling is unlikely at high overpotential region and proton transfer mechanism should be govern by the over-the-barrier path.

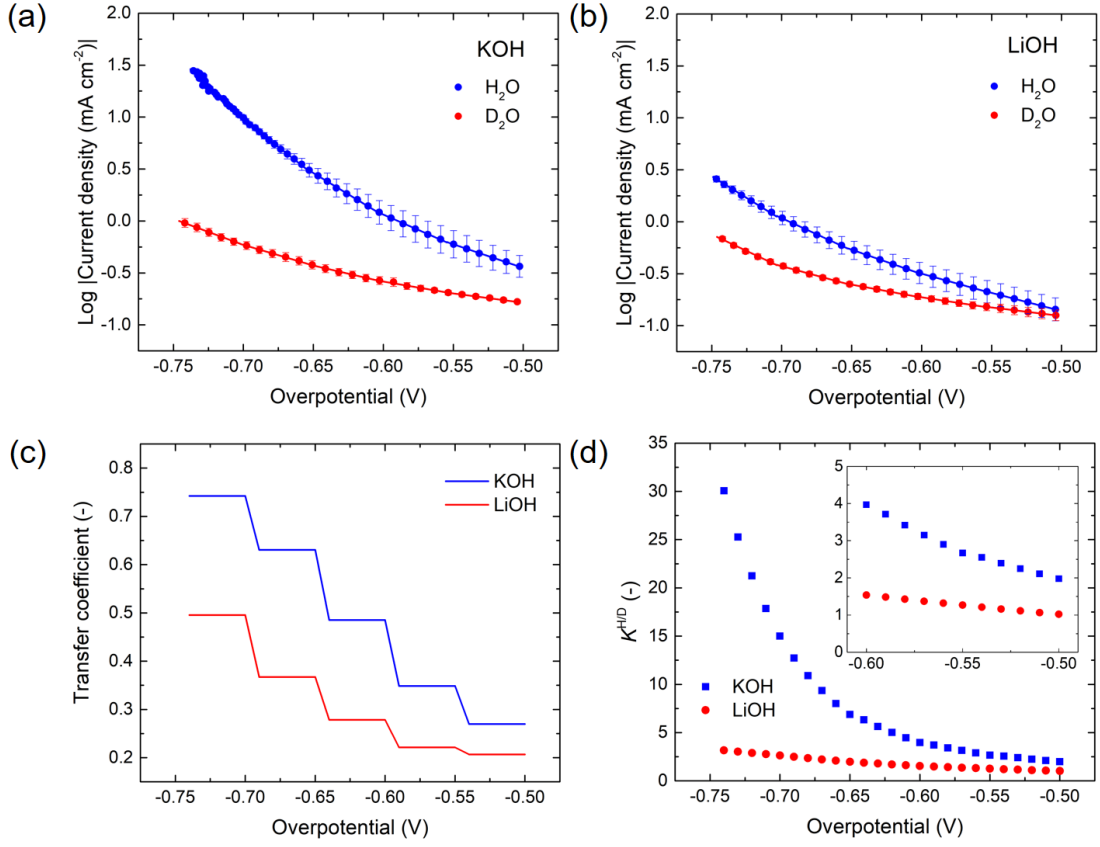

Figure 2: Tafel plots for the hydrogen evolution on Au(111) in (a) 0.1 M KOH in $H_2O$ and $D_2O$ and (b) 0.1 M LiOH in $H_2O$ and $D_2O$. The potential scanning rate is 0.050 V $s^{-1}$. The linear fitting of the plots is conducted for every 0.05 V. (c) Transfer coefficient ($\alpha$) for the $H_2O$ systems and (d) $K^{H/D}$ vs overpotential diagram for the HER on Au(111) in 0.1 M KOH and LiOH. The inset shows the enlarge diagram at low overpotential region. $\alpha$ and $K^{H/D}$ values are calculated for every 0.05 and 0.01 V, respectively.

This anomalous behavior in QCT can be interpreted by the theory of nonadiabatic processes for the Volmer step at the electrified interface, which describes quantum mechanical transition of the vibronic states as a key of electrochemical processes [23, 45]. The free energy diagrams of the vibronic states can be represented as the parabolas [23, 46], as shown in Fig. 3. The nonadiabatic processes occur at intersection points ($X^*$) between reactant and product vibronic states. A overlap of reactant and product vibronic states are defined by $V_{ak} = \langle \mathbf{a}|H_e|\mathbf{k}\rangle$, where $\langle \mathbf{a}|$, $|\mathbf{k}\rangle$, and $H_e$ are reactant electronic states, product electronic states, and electronic Hamiltonians, respectively

[23, 47]. The previous study suggested that the adiabaticity and nonadiabaticity for the Volmer step are influenced by the water structure at electrified solid-liquid interface [23]. Based on this theory of nonadiabaticity in proton-coupled electron transfer, the cation-dependent anomalous QCT mechanism in the HER on Au (111) is suggested to be the result from the water structure induced by cations, which is indeed an example of the observation of nonadiabaticity at electrified solid-liquid interfaces.

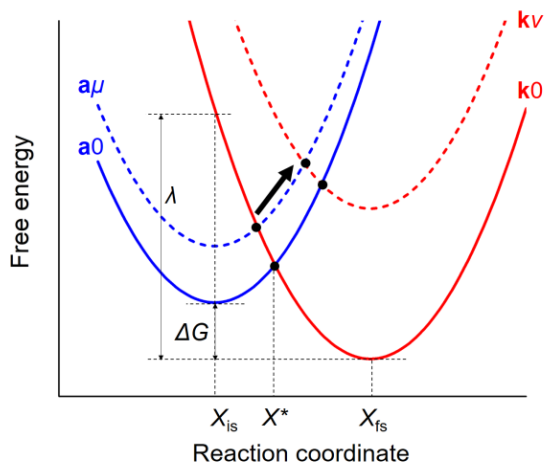

Figure 3: Free energy curves for products (blue) and reactants (red) related by the reorganization energy ($\lambda$) and the Gibbs free energy change ($\Delta G$) under the framework of Soudackov-Hammes-Schiffer theory [8]. $X_{is}$ and $X_{fs}$ represent the reaction coordinates of initial and final states, respectively. The solid and dashed curves show the ground and excited vibronic states, respectively. The circles represent the intersection points ($X^*$) between reactant and product vibronic states. Nonadiabatic transitions occur at the intersection points between excited reactant and product vibronic states (**a**$\mu$ → **k**$\nu$) represented by the black bold arrow.

This mechanism to explain nonadiabaticity for the HER by involves the water dimer adsorbed on the surface, as shown in Fig. 1. However, this mechanism cannot fully picture this observed PCET electrode processes, especially in the case of the LiOH condition because no quantum process was suggested in this system (Fig. 2d). Two of the main lacking factors in the previous theory are (1) potential-dependent water dynamics at electrode surface and (2) cation effects. First for potential dependent water dynamics, at low $\eta$ region, because the negative charge of electrode surface is weak, the distance between reactant water and electrode is known to be too far to trigger proton tunneling as indicated by $\beta$ which shows the position of a reaction intermediate between the proton donor ($H_2O_{ad}$) and the electrode surface as the proton acceptor (Fig. 2(c), note $\alpha = \beta$ in this study). Therefore, the RDS would be the long-range electron transfer from the electrode to $H_2O_{ad}$ ($H_2O + e^- \rightarrow H_2O^-$) at low $\eta$ region, which is consistent with the $K^{H/D} = 1$ in the LiOH condition suggesting no proton transfer at the RDS [26]. On the other hand, in the KOH condition, the semi-classical proton transfer is considered as the RDS at low $\eta$ region by the $K^{H/D}$ values ($2 < K^{H/D} < 10$), which indicates that the distance between reactant water and electrode in the KOH condition is closer than that in the LiOH condition as the PCET was observed at this condition. In general, quantum proton tunneling requires a narrow distance for the adsorption of water molecules on the surface with the H-down orientation [26]. Therefore cation effect is suggested play an important role to trigger the nonadiabaticity. Cations is known to influence the potential-dependent water structure by electrostatic interactions. Therefore, the cation-induced water structure is also suggested to be a pivotal factor for the potential-dependent PCET mechanism.

Aiming to give a comprehensive microscopic interpretation to the observation on the electrode process in the KOH and LiOH conditions, we propose the nonadiabatic cation-dependent PCET mechanism based on the electrified interfacial structure. Applied potential is well known to play an important role in the kinetics for the PCET mechanism by tuning the reaction barrier height and width. In addition, the orientation and the hydrogen-bonding structure of the water molecules are presumed to be the key factors determining the proton transfer kinetics and mechanism. The orientation of the water molecules is shifted between the H-up and H-down around the potential of zero charge (pzc) of an electrode surface [48−50]. The PCET electrode process in Step II proceeds between two water molecules adsorbed on Au (Fig. 1). As suggested by the previous theory, the nonadiabatic PCET requires the specific dimer structure of H-down and H-up water molecules coordinated with each other by a hydrogen bonding [23]. However, the water molecules prefers H-down orientation at high HER overpotentials. Here the cation effect is the key why the KOH system shows nonadiabatic PCET and the LiOH not. As $K^+$ can destabilize the rigid hydrogen-bonding structure, this feature of $K^+$ leads to the specific orientation of the water dimer configuration which triggers the nonadiabatic PCET (Fig. 4. left). In contrast, $Li^+$ strongly interacts with the water molecules by non-covalent interactions [51]. These strong interactions between $Li^+$ and the water molecules will lead to a rigid water structure therefore the vast majority of $H_2O$ at the electrode surface will be fixed to H-down orientation which cannot induce nonadiabatic PCET (Fig. 4 right). On the contrary, $K^+$ weakly interacts with the water molecules and therefore is able to move at the solid-liquid interface. This feature allows $K^+$ to approach to the electrode surface to break the interfacial water structure at high HER overpotentials [39, 51], which can alter the interfacial charge associated with the water orientation. Thus the weak interaction between the $K^+$ and the water molecules is suggested to trigger to form the specific water dimer configuration facilitating the nonadiabaticity during the PCET as indicated by the theory [23].

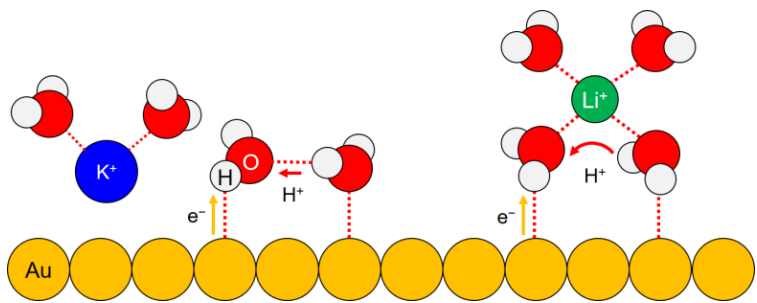

Figure 4: Schematic model of the PCET electrode process for the adsorption of hydrogen atoms during the HER on Au(111) in KOH and LiOH solutions at higher overpotential region. $K^+$ weakly interacts with the water molecules and therefore is able to approach to the electrode surface to break the interfacial water structure leading to the favorable water dimer configuration to trigger nonadiabatic PCET (left). Besides, the water structure organized by noncovalent interactions with $Li^+$ lead to proton transfer without quantum effects (right).

In summary, we observed a nonadiabatic cation-dependent PCET during potential-induced hydrogen evolution reaction on the Au(111) in KOH system. The anomalous QCT attributed to the manifestation of nonadiabaticity. The weak interaction between $K^+$ and interfacial water molecules could proceed the formation of the specific water dimer structure facilitating this quantum effect, while $Li^+$ cation can strongly coordinate with water molecules and form a rigid interfacial water structure by non-covalent interactions toward inhibiting a manifestation of nonadaibaticity.

Acknowledgments

K.S. and T.K. are indebted to the National Institute for Materials Science. This work was financially supported by JSPS KAKENHI Grant Number 21K18941(for KS) and 21J00688 (for TK), Japan.